\documentstyle[12pt] {article}
\def\zid{1\kern-0.36em\llap~1}

\newcommand{\beq}{\begin{equation}}
\newcommand{\ber}{\begin{eqnarray}}
\newcommand{\eeq}{\end{equation}}
\newcommand{\eer}{\end{eqnarray}}

\begin{document}

\begin{titlepage}
\rightline{SUNY BING 7/2/98}
\rightline{hep-th/9807048}
\vspace{1mm}
\begin{center}
{\bf  EIGENSTATES OF PARAPARTICLE CREATION OPERATORS}\\
\vspace{2mm}
Sicong Jing\footnote{On leave from:  Department of
Modern Physics, University of Science and Technology of 
China, Hefei, 230026, P.R. China.} and 
Charles A. Nelson\footnote{Electronic address: cnelson @ 
bingvmb.cc.binghamton.edu }\\
{\it Department of Physics, State University of New York at 
Binghamton\\
Binghamton, N.Y. 13902-6016}\\[2mm]
\end{center}


\begin{abstract} 
Eigenstates of the parabose and parafermi creation operators 
are constructed.  In the Dirac contour representation, the 
parabose eigenstates correspond to the dual vectors of the  
parabose coherent states.  In order $p=2$, conserved-charge 
parabose creation operator eigenstates are also constructed.  
The contour forms of the associated resolutions of unity are 
obtained. 
\end{abstract}

\end{titlepage}

\section{Introduction}

Eigenstates of the ordinary bose creation operator were 
constructed in Ref. [1] using Heitler's contour integral form 
of the $\delta$-function[2].  These eigenstates   
correspond to the dual vectors of the coherent states in 
Dirac's contour representation of boson systems[3,1,4]. It is 
natural to investigate whether such creation operator 
eigenstates can also be constructed for paraparticles[5,6].  
Parabose coherent states were proposed in [7], parafermi 
coherent states in [6], and recently, parabose squeezed 
states in [8].  

In this paper in Sec. 2, the eigenstates for the parabose 
creation operator are constructed.  Heitler's form of the 
$\delta$-function is used, so the expansion coefficients for 
these eigenstates in the parabose number basis are actually 
distributions.  In Sec. 3, paragrassman numbers are used in 
the construction of the eigenstates of the parafermi creation 
operator.  In the number basis, the expansion coefficents for 
the $f$ eigenstates and for the $f^{\dagger}$ 
eigenstates are paragrassman numbers, and so in this case, 
there is also an enlargement of the usual Hilbert space 
description. Lastly, in Sec. 4, the conserved-charge parabose 
creation operator eigenstates are constructed for the 
two-mode parabose system in order $p=2$.  In each section, 
the respective contour forms of the resolution of unity are 
derived.   
 
\section{Eigenstates of the Parabose $a^{\dagger}$ Operator}

For a single-mode parabose system, the number basis is 
\begin{equation}
|n>=\frac{(a^{\dagger })^n}{\sqrt{[n]!}}|0>,\;\;N_B|n>=n|n>, 
\end{equation}
where $N_B$ is the number operator $N_B=\frac 12\{a^{\dagger 
},a\}-\frac p2$ with $p$ the order of the parastatistics.  
The eigenvalue of the deformed parabose number operator $[N_B 
]$ is  
\begin{equation}
[n]=n+\frac{p-1}2(1-(-)^n),
\end{equation}
with $[n]!=[n][n-1]\cdots 
[1],\;\;[0]!\equiv 1
 $.
The parabose number states statisfy
\begin{equation}
a|n>=\sqrt{[n]}|n-1>,\;\;a^{\dagger }|n>=\sqrt{[n+1]}|n+1>
\end{equation}
In this basis, the unnormalized coherent states [7] are
\begin{equation}
|z>=\sum_{n=0}^\infty 
\frac{z^n}{\sqrt{[n]!}}|n>=E(za^{\dagger
})|0>,\;\;E(x)\equiv \sum_{n=0}^\infty \frac{x^n}{[n]!}
\end{equation}
with $a|z>=z|z>$.

We denote the eigenstate of the creation operator $a^{\dagger 
}$ by a
``primed'' ket  $|z>^{^{\prime }}$,
\begin{equation}
a^{\dagger } |z>^{^{\prime }}=z^{*}|z>^{^{\prime }},
\end{equation}
and expand it  $|z>^{^{\prime 
}}=\sum_{n=0}^\infty c_n(z^{*})|n> $ in the number basis.  
By (3), the resulting recursion relations are
\begin{equation}
c_0\;z^{*}=0,\;c_1\;z^{*}=\sqrt{[1]}\;c_0,\cdots 
;\;c_n\;z^{*}=\sqrt{[n]}\;c_{n-1},
\end{equation}
or $c_n=\frac{ \sqrt{[n]!} }{ (z^{*})^n }c_0$. 
By the Cauchy integral formula 
for an analytic function 
$f(z^{*})$, or alternatively by
use of Heitler's $\delta $-function [4]  in the contour 
integral\footnote{Note $
f(0)=\oint_{C^{*}}\frac{dz^{*}}{2\pi i}\;f(z^{*})\delta 
(z^{*})$; the n-th
derivative $\;f^{(n)}(0)=n!\;\oint_{C^{*}}\frac{dz^{*}}{2\pi 
i}\;\frac{%
f(z^{*})}{(z^{*})^{n+1}}=(-)^n\int_{-\infty }^\infty 
\;dx\;f(x)\;\delta
^{(n)}(x)$.  } , it follows[1] that 
\begin{equation}
c_0=\frac 1{z^{*}}|_{C^{*}}=\delta (z^{*})
\end{equation}
\begin{equation}
c_n=\frac{\sqrt{[n]!}}{(z^{*})^{n+1}}|_{C^{*}}=\frac{(-
)^n\sqrt{[n]!}}{n!%
}\;\delta ^{(n)}(z^{*}).
\end{equation}
The notation $|_{C^{*}}$ means that the subsequent 
integration over $%
z^{*}$ must be over the counterclockwise contour $C^{*}$ 
enclosing the 
origin in the
complex $z^{*}$ plane.

So, the eigenstate of $a^{\dagger }$ is 
\begin{equation}
|z>^{'}=\sum_{n=0}^\infty \frac{(-
)^n\sqrt{[n]!}}{n!}\;\delta
^{(n)}(z^{*})|n>=\sum_{n=0}^\infty 
\frac{\sqrt{[n]!}}{(z^{*})^{n+1}}%
|n>|_{C^{*}}=\sum_{n=0}^\infty \frac{(a^{\dagger 
})^n}{(z^{*})^{n+1}}%
|0>|_{C^{*},}
\end{equation}
and, formally (in the number basis)
\begin{equation}
|z>^{'}=\frac 1{z^{*}-a^{\dagger }}|0>|_{C^{*}}.
\end{equation}
Note that the action of integer powers of $(a^{\dagger })^m$ 
removes the
contribution of the number states $n<m$ in this expansion; 
e.g. 
$$
a^{\dagger }|z>^{'}=a^{\dagger }\;\frac 1{z^{*}-a^{\dagger
}}|0>|_{C^{*}}=z^{*}\;\frac 1{z^{*}-a^{\dagger
}}|0>|_{C^{*}}-|0>|_{C^{*}}=z^{*}\;|z>^{'} 
$$
since the $|0>|_{C^{*}}$ term gives no contribution because 
of the
subsequent contour integration.

In the Dirac contour representation, the dual vector $<\alpha 
|$ of the parabose conherent state $|\alpha >$, given above 
in (4), is [10]
\begin{equation}
<\alpha |=\sum_{n=0}^\infty <n|\frac{(\alpha 
^{*})^n}{\sqrt{[n]!}}%
\;\rightarrow \sum_{n=0}^\infty 
\frac{\sqrt{[n]!}}{z^{n+1}}\frac{(\alpha
^{*})^n}{\sqrt{[n]!}}=\frac 1{z-\alpha ^{*}}, \; \; 
( |z| > |\alpha|).
\end{equation}
with $<\alpha |a^{\dagger} =<\alpha | \alpha^{*}$.  Thus, the 
eigenstate $|z>^{'}$ of $a^{\dagger}$ in the number basis 
corresponds to the parabose coherent state's eigenbra 
$<\alpha |$ of $a^{\dagger}$ in the Dirac contour 
representation. 

The inner product of the unnormalized 
parabose coherent state $%
|w>$ and the  
eigenstate  $|z>^{'}$ is given by 
$$
<w|z>^{'} =\sum_{n,m=0}^\infty <n|
\frac{(w^{*})^n}{\sqrt{[n]!}}\frac{\sqrt{[m]!}}{(z^{*})^{
m+1}}%
|m>|_{C^{*}}=\sum_{n=0}^\infty 
\frac{(w^{*})^n}{(z^{*})^{m+1}}|_{C^{*}} 
$$
\begin{equation} 
=\frac 1{z^{*}-w^{*}}|_{C^{*}}=\delta (z^{*}-
w^{*}),\;\;(|z|>|w|)
\end{equation}
and so they satisfy the ``contour form" of the resolution of 
unity, see [3],  
\begin{equation}
\oint_{C^{*}}\frac{dz^{*}}{2\pi 
i}\;|z>^{'} <z|=\sum_{n,m=0}^\infty
\oint_{C^{*}}\frac{dz^{*}}{2\pi 
i}\;\frac{\sqrt{[n]!}}{(z^{*})^{n+1}}|n><m|%
\frac{(z^{*})^m}{\sqrt{[m]!}}=\sum_{n=0}^\infty |n><n|=I.
\end{equation}

{\bf Remark:}  This resolution of unity can be used to derive 
a contour 
integral
expressions for the parabose Hermite polynomials[11]: From 
(13), the parabose
coordinate eigenstate $|x>$ can be written as 
\begin{equation}
|x>=\sum_{n=0}^\infty |n>\frac{\sqrt{[n]!}}{2\pi i 
}\oint_{C^{*}}\frac{%
dz^{*}}{(z^{*})^{n+1}}<z|x>,
\end{equation}
where $<x|z>$ is the wave function of the parabose coherent 
state in the
parabose coordinate representation. We consider
\begin{equation}
<x|z>=\frac 1x<x|\hat x|z>=\frac 1{x\sqrt{2}}<x|(a+a^{\dagger 
})|z>
\end{equation}
From(3), c.f. eq.(14) for the parabose deformed derivative  
$D/Dz$ in [10],   
\begin{equation}
a^{\dagger }|z>=\frac \partial {\partial z}|z>+ ( \frac{p-
1}{2z} ) |z>-( \frac{p-1}{%
2z} ) |-z>,
\end{equation}
so (15) gives 
\begin{equation}
\frac \partial {\partial z}<x|z>=(-z+x\sqrt{2}-\frac{p-
1}{2z})<x|z>+ ( \frac{p-1%
}{2z} ) <x|-z>.
\end{equation}
This has the solution 
\begin{equation}
<x|z>=N_0e^{-\frac{z^2}2-\frac{x^2}2}E(\sqrt{2}xz)
\end{equation}
with $N_0$ a normalization constant. Substituting this into 
(14) gives
\begin{equation}
|x>=N_0e^{-\frac{x^2}2}\sum_{n=0}^\infty 
|n>\frac{\sqrt{[n]!}}{2\pi i }%
\oint_{C^{*}}\frac{dz^{*}}{(z^{*})^{n+1}}e^{-
\frac{(z^{*})^2}2}E(\sqrt{2}%
xz^{*})
\end{equation}
But from [11], in the parabose coordinate representation
\begin{equation}
|x>=\sum_{n=0}^\infty |n><n|x>=N_0e^{-
\frac{x^2}2}\sum_{n=0}^\infty |n>\frac{%
H_n^{(p)}(x)}{\sqrt{2^n[n]!}},
\end{equation}
\begin{equation}
H_n^{(p)}(x)=[n]!\sum_{k=0}^{[\frac n2]^{^{\prime }}}\frac{(-
)^k(2x)^{n-2k}}{%
k![n-2k]!},
\end{equation}
where $[k]^{^{\prime }}$ denotes the largest integer less 
than or equal to $k
$. So since the $|n>$ are complete, 
\begin{equation}
H_n^{(p)}(x)=\frac{[n]!}{2\pi i  
}\oint_C\frac{dz}{(z)^{n+1}}e^{-\frac{%
(z)^2}2}E(\sqrt{2}xz).
\end{equation}

\section{Eigenstates of the Parafermi $f^{\dagger}$ Operator}

In the finite dimensional Hilbert space of a single-mode 
parafermi system, the number states can be 
written as
\begin{equation}
|n>=\frac{(f^{\dagger })^n}{\sqrt{\{n\}!}}|0>,\;\;N_f|n>=n|n>
\end{equation}
where $N_f=\frac 12[f^{\dagger },f]+\frac p2$ is the 
parafermi number
operator.  Here  
\begin{equation}
\{n\}=n(p+1-n),\;\;\{n\}!=\{n\}\{n-1\}\cdots 
\{1\},\;\;\{0\}!\equiv 1.
\end{equation}
with $n$ an integer, $0\leq n\leq p$. In this basis,\ 
$f^{\dagger }|n>=\sqrt{%
\{n+1\}}|n>,\;\;f|n>=\sqrt{\{n\}}|n-1>$. Since\ $f^{\dagger 
}|p>=0$, there is the useful fact that 
\begin{equation}
|n>=\sqrt{\frac{\{n\}!}{\{p\}!}}\;\;f^{p-n}\;|p>.
\end{equation}

To describe the parafermi eigenstates of $f$ ( 
and of $%
f^{\dagger }$ ) in this number basis, we use [6] paragrassman 
numbers $\xi $
obeying $\xi ^{p+1}=0.$ The unnormalized eigenstate of the 
parafermi
annihilation operator $f$ 
\begin{equation}
|\xi >=\sum_{n=0}^p|n>\frac{\xi ^n}{\sqrt{\{n\}!}}
\end{equation}
satisfies the eigenequation $f|\xi >=|\xi >\xi $. In this 
formulation, $|\xi >$ is expandable in the number basis, c.f. 
[6]. Note that 
$\xi $ stands to
the right of $|\xi >$. The overlap of two eigenstates $|\xi 
>$ and $|\zeta >$
is 
\begin{equation}
<\xi |\zeta >=\sum_{n=0}^p\frac{(\xi ^{*})^n\zeta ^n}{\{n\}!}
\end{equation}
where $\xi ^{*}$ is the conjugate of $\xi .$ By the 
paragrassmann integral
formula (see appendix)
\begin{equation}
\int \xi ^n\;d\mu (\xi ,\xi ^{*})\;(\xi ^{*})^m=\delta 
_{n,m}\;\{n\}!
\end{equation}
and (26), there is the resolution of unity 
$$
\int |\xi >d\mu (\xi ,\xi ^{*})<\xi |=I,
$$
\begin{equation} 
d\mu (\xi ,\xi ^{*})=d^p\xi ^{*}\;d^p\xi \;e^{-\frac 
12[\xi ^{*},\xi ]}.
\end{equation}

To constuct the eigenstates of the parafermi creation 
operator $f^{\dagger }$%
, we recall (25) and consider
\begin{equation}
|\xi >^{^{\prime }}=\sum_{n=0}^p|n>(-\xi ^{*})^{p-
n}\sqrt{\frac{\{n\}!}{%
\{p\}!}}.
\end{equation}
These are the desired eigenstates since 
$$
f^{\dagger }|\xi >^{^{\prime }}=\sum_{n=0}^{p-1}|n+1>(-\xi 
^{*})^{p-n}\sqrt{%
\frac{\{n+1\}!}{\{p\}!}}=\sum_{n=1}^p|n>(-\xi ^{*})^{p-
n+1}\sqrt{\frac{\{n\}!%
}{\{p\}!}} 
$$
\begin{equation}
=-|\xi >^{^{\prime }}\;\;\xi ^{*}.
\end{equation}
The overlap of these eigenstates is
\begin{equation}
^{^{\prime }}<\xi |\zeta >^{^{\prime 
}}=\sum_{n=0}^p\frac{\{n\}!}{\{p\}!}\xi
^{p-n}(\zeta ^{*})^{p-n}=\sum_{n=0}^p\frac{\{p-
n\}!}{\{p\}!}\xi ^n(\zeta
^{*})^n
\end{equation}
As for the $f$ eigenstates in (29), the eigenstates $|\xi 
>^{^{\prime 
}}$ obey a 
resolution of unity 
\begin{equation}
\int |\xi >^{^{\prime }}d\mu (\xi ^{*},\xi ) \;\; ^{^{\prime 
}}<\xi |=I
\end{equation}
where 
\begin{equation}
d\mu (\xi ^{*},\xi )=d^p\xi \;d^p\xi ^{*}\;e^{-\frac 
12[\xi ,\xi ^{*}]}.
\end{equation}
This follows from (26) and (30) by 
$$
\int |\xi >^{^{\prime }}d\mu (\xi ^{*},\xi ) \;\; ^{^{\prime 
}}<\xi | 
$$
$$
=\sum_{n,m=0}^p|n>\int (-\xi ^{*})^{p-n}\;d\mu (\xi 
^{*},\xi ) (-\xi )^{p-m}<m|\;%
\frac{\sqrt{\{n\}!\{m\}!}}{\{p\}!} 
$$
\begin{equation}
=\sum_{n,m=0}^p\frac{\{n\}!}{\{p\}!}\{p-n\}!|n><n|=I
\end{equation}
where we have used the fact that $\{n\}!=\frac{n!\;p!}{(p-
n)!}.$

Furthermore, with the aid of the differentiation formula (see 
appendix)
\begin{equation}
\frac \partial {\partial \xi }\;\xi ^n=\{n\}\;\xi ^{n-1}=\xi 
^n\;\frac{%
\overleftarrow{\partial }}{\partial \xi },\;\;(0\leq n\leq 
p),
\end{equation}
we have 
\begin{equation}
f^{\dagger }\;|\xi >=|\xi >\;\frac{\overleftarrow{\partial 
}}{\partial \xi }%
,\;\;\;f\;|\xi >^{^{\prime }}=-\;|\xi >^{^{\prime 
}}\;\frac{\overleftarrow{%
\partial }}{\partial \xi ^{*}},
\end{equation}
which give the matrix elements of $f^{\dagger}$, $f$,%
$$
<\xi |\;f^{\dagger }\;|\xi >^{^{\prime }}=\xi ^{*}\;<\xi |\xi 
>^{^{\prime
}}=-\;<\xi |\xi >^{^{\prime }}\;\xi ^{*}, 
$$
\begin{equation}
<\xi |\;f\;|\xi >^{^{\prime }}=\frac \partial {\partial \xi 
^{*}}\;<\xi |\xi
>^{^{\prime }}=-\;<\xi |\xi >^{^{\prime 
}}\;\frac{\overleftarrow{\partial }}{%
\partial \xi ^{*}}.
\end{equation}
Alternatively, these equations follow from (26) and (30)  
since 
\begin{equation}
<\xi |\xi >^{^{\prime }}=\frac{(-
)^p}{\sqrt{\{p\}!}}\sum_{n=0}^p(-)^n\;(\xi
^{*})^n(\xi ^{*})^{p-n}
\end{equation}

Lastly, as in the parabose case (13), there is a 
contour-like-form resolution of unity for the $f^{\dagger}$ 
and $f$ eigenstates: %
$$
\int |\xi >^{^{\prime }}\;d^p\xi ^{*}\;<\xi | 
=\sum_{n,m=0}^p\sqrt{\frac{\{n\}!}{\{p\}!}}|n>\int (-\xi 
^{*})^{p-n}\;d^p\xi
^{*}\;\;(\xi ^{*})^m<m|\;\frac 1{\sqrt{\{m\}!}} 
$$
\begin{equation}
=\sum_{n=0}^p|n><n|=I
\end{equation}
where we have used 
\begin{equation}
\int \;d^p\xi ^{*}\;(\xi 
^{*})^p=p!,\;\;\int \;d^p\xi
^{*}\;(\xi ^{*})^n=0\;\;(0\leq n\leq p),
\end{equation}
and $\xi 
^{*}\;d^p\xi ^{*}=-d^p\xi
^{*}\;\xi ^{*}$.  Note that in (40), as in the parabose case 
(13), the integration is only over a single variable in the 
contour form of the resolution of unity, whereas in (29) and 
(35) it is over two variables as for the usual parabose 
coherent states. 

\section{Conserved-Charge Parabose Creation Operator \newline 
Eigenstates 
for Order $p=2$}

The parabose creation and annihilation operators for the 
two-mode system satisfy the trilinear commutation relations 
\begin{equation}
\begin{array}{c}
\lbrack a_k,\{a_l^{\dagger },a_m\}]=2\delta 
_{kl}a_m, \; [a_k,\{a_l^{\dagger
},a_m^{\dagger }\}]=2\delta _{kl}a_m^{\dagger }+2\delta 
_{km}a_l^{\dagger },
\\ 
\lbrack a_k,\{a_l,a_m\}]=0, \hspace{2pc} (k,l,m=1,2)
\end{array}
\end{equation}
where $a_1=a$, $a_2=b$.  Since $ab \neq ba$ for $p \geq 2$,   
there is a degeneracy in the states with $n$ parabosons $a$ 
and $m$ parabosons $b$.  For such states, we find [9] that 
the degree of degeneracy is ``$\min (n,m)+1$".  The complete 
set 
of state vectors is\footnote{Note that here in Sec. 4, but 
not in 
Sec. 2,  $[x]$ denotes the 
integer 
part of $x$ for $x\geq 0$.} 
\begin{equation}
|n,m;i>=\frac 1{\sqrt{N_{n,m}^i}}(a^{\dagger })^{n-
i+S}(b^{\dagger })^{m-2[%
\frac{i-S}2]}(a^{\dagger }b^{\dagger })^{2 [\frac{i-
S}2]}(a^{\dagger })^{i-S-2[%
\frac{i-S}2]}|0> 
\end{equation}
where $N_{n,m}^i$ is the normalization constant, and 
$S={ \frac 12 }(1-(-)^m)$, and $i$ is the degeneracy index $ 
1\leq i\leq \min (n,m)+1$.  
For parastatistics of order $p=2$, the $\{ |n,m;i> \}$ are an 
orthonormal set basis vectors with normalization constant 
\begin{equation}
(N_{n,m}^i)^2=2^{n+m}[\frac{n+i}2]![\frac{n+1-
i}2]![\frac{m+i}2]![\frac{m+1-i%
}2]!
\end{equation}
In this basis, $a^{\dagger}, b^{\dagger}, a, b$ also act as 
raising and lowering operators (the explict fomulas are given 
in eqs.(15-18) in [9]).

If we consider $a$ and $b$ to be two types of parabose quanta 
possessing abelian charges ``+1" and ``-1", then the charge 
operator is  
\begin{equation}
Q\equiv N_a-N_b
\end{equation}
with $ N_a=\frac 12\{a^{\dagger },a\}-
1,\; N_b=\frac
12\{b^{\dagger },b\}-1  $.  This charge operator $Q$ commutes 
with the operators $a^{\dagger }b^{\dagger }$\ 
and 
$b^{\dagger
}a^{\dagger },$ so their common eigenstate should satisfy 
the 
eigenequations
\begin{equation}
\begin{array}{c}
Q\;|\;q,z,w>^{^{\prime }}=q|\;q,z,w>^{^{\prime }}, \\ 
\;a^{\dagger }b^{\dagger
}|\;q,z,w>^{^{\prime }}=w^{*}|\;q,z,w>^{^{\prime 
}},\;b^{\dagger 
}a^{\dagger
}|\;q,z,w>^{^{\prime }}=z^{*}|\;q,z,w>^{^{\prime }}.
\end{array}
\end{equation}
Expanding $|\;q,z,w>^{^{\prime }}$ in terms of the complete 
set of
orthonormal basis vectors $|n,m;i>$ for the two-mode parabose 
system, for $q\geq 0$ we have from the $Q$ 
eigenequation (46) 
\begin{equation}
|\;q,z,w>^{^{\prime }}=\sum_{m=0}^\infty 
\sum_{i=1}^{m+1}c_{q+m,m}^i\;|q+m,m;i>
\end{equation}
From the remaining two eigenequations, we obtain the 
coefficients
$$
c_{q+m,m}^i=
\frac{(-)^m2^m\sqrt{[\frac{q+m+i}2]![\frac{q+m+1-
i}2]!}}{\sqrt{[\frac q2]![%
\frac{q+1}2]![\frac{m+i}2]![\frac{m+1-i}2]!}}\delta ^{\left( 
s\right)
}(w^{*})\delta ^{\left( r\right) }(z^{*})
$$ 
\begin{equation}
=\frac 
1{\sqrt{[\frac q2]![%
\frac{q+1}2]!}}\frac{2^m\sqrt{[\frac{m+i}2]![\frac{m+1-
i}2]![\frac{q+m+i}2]![%
\frac{q+m+1-
i}2]!}}{(w^{*})^{1+s}(z^{*})^{1+r}}\;|_{C^{*},B^{*
}}
\end{equation}
where the integers 
$$
\begin{array}{c}
r\equiv \left[ 
\frac{m-(-)^{q+m+i}i}2+\frac{1-(-)^q}4\right],  \\ s\equiv 
\left[ \frac{%
m+(-)^{q+m+i}i}2+\frac{1+(-)^q}4\right] 
\end{array}
$$
The counterclockwise contours $C^{*}$ and $B^{*}$ enclose 
respectively the
origins in the complex $z^{*}$ and $w^{*}$ planes. Since for 
a specific $q$%
-sector the overall $\frac 1{\sqrt{[\frac 
q2]![\frac{q+1}2]!}}$ is constant,
we omit it in the following analysis.

We  list results for only the $q\geq 0$ sector: in it the 
unnormalized dual vectors 
are 
\begin{equation}
|\;q,z,w>^{^{\prime }}=\sum_{m=0}^\infty 
\sum_{i=1}^{m+1}\frac{2^m\sqrt{[\frac{m+i%
}2]![\frac{m+1-i}2]![\frac{q+m+i}2]![\frac{q+m+1-i}2]!}}{%
(w^{*})^{1+s}(z^{*})^{1+r}}|q+m,m;i>\;|_{C^{*},B^{*}},
\end{equation}
whereas the unnormalized parabose conserved-charge coherent 
states 
themselves are[9]  
\begin{equation}
|\;q,v,u>\ =\sum_{m=0}^\infty 
\sum_{i=1}^{m+1}\frac{v^ru^s}{2^m \sqrt{[%
\frac{m+i}2]![\frac{m+1-i}2]![\frac{q+m+i}2]![\frac{q+m+1-
i}2]!}}|q+m,m;i>.
\end{equation}
The inner product of 
$|\;q,z,w>^{^{\prime }}$ and 
$|\;q,v,u>$ is 
$$
<\;q,v,u|\;q,z,w>^{^{\prime }}=\sum_{m=0}^\infty 
\sum_{i=1}^{m+1}\left( 
\frac{v^{*}}{z{}^{*}}\right) ^r\left( 
\frac{u^{*}}{w^{*}}\right)
^s\left( \frac 1{z{}^{*}w^{*}}\right) \;|_{C^{*},B^{*}} 
$$
\begin{equation} 
=\frac
1{z{}^{*}-v{}^{*}}|_{C^{*}}\frac 1{w{}^{*}-
u{}^{*}}|_{B^{*}}=\delta
(z{}^{*}-v{}^{*})\;\delta
(w{}^{*}-
u{}^{*}),\;\;(|z^{*}|>|v^{*}|,|w^{*}|>|u^{*}|).
\end{equation}
These satisfy the contour form of the resolution of unity%
$$
\oint_{C^{*}}\oint_{B^{*}}\frac{d\;z^{*}}{2\pi 
i}\frac{d\;w^{*}}{2\pi i}%
\;|\;q,z,w>^{^{\prime }}<\;q,z,w| 
$$
\begin{equation}
=\sum_{m=0}^\infty \sum_{i=1}^{m+1}|q+m,m;i><q+m,m;i|=I_q.
\end{equation}
where $I_q$ is the unity operator in the $q \geq 0 $ sector.

In summary, working in the number basis, in this paper we 
construct the creation operator eigenvectors for single-mode 
parabosons and parafermions, and for the two-mode 
conserved-charge parabosons.  The contour forms of the 
associated resolutions of unity are obtained.

This work was partially supported by the National Natural 
Science Foundation of China and by U.S. Dept. of Energy 
Contract No. DE-FG 02-96ER40291.

\newpage 

\begin{center}
{\bf Appendix}
\end{center}

\appendix

\section{Proof of paragrassman integration formula:}

We write $\xi =\sum_{i=1}^p\;\xi _i$, 
where the Green components $\xi _i$ satisfy the relations
\begin{equation}
\{\xi _i,\xi _i\}=0,\;\;[\xi _i,\xi _j]=0\;\;(i\neq j).
\end{equation}
Also, 
$$
\frac 12[\xi ^{*},\xi ]=\sum_{i=1}^p\;\xi _i^{*}\xi 
_i,\;(\frac 12[\xi
^{*},\xi ])^2=2!\;\sum_{i<j}\xi _i^{*}\xi _j^{*}\xi _i\xi 
_j,\;\cdots,  
$$
\begin{equation}
(\frac 12[\xi ^{*},\xi ])^n=n!\;\sum_{i_1<\cdots <i_n}\xi 
_{i_1}^{*}\cdots
\xi _{i_n}^{*}\xi _{i_1}\cdots \xi _{i_n},\;\cdots ,(\frac 
12[\xi ^{*},\xi
])^p=p!\;\xi _1^{*}\cdots \xi _p^{*}\xi _1\cdots \xi _p,
\end{equation}
so%
$$
e^{-\frac 12[\xi ^{*},\xi ]}=1+\sum_i\xi _i\xi _i^{*} 
$$
\begin{equation}
+\sum_{i<j}\xi _i\xi _j\xi _i^{*}\xi 
_j^{*}+\cdots
+\sum_{i_1<\cdots <i_n}\xi _{i_1}\cdots \xi _{i_n}\xi 
_{i_1}^{*}\cdots \xi
_{i_n}^{*}+\cdots+\xi _1\cdots \xi _p\xi _1^{*}\cdots \xi 
_p^{*}
\end{equation}
For paragrassman integration, we adopt
\begin{equation}
\int d^p\xi \;\xi _1\cdots \xi _p=1,\;\int d^p\xi \;\xi 
_{i_1}\cdots \xi
_{i_n}=0\;(0\leq n<p)
\end{equation}
where $d^p\xi \equiv d\xi _1\cdots d\xi _p$.  These give 
(41).

By (56), the integral $\int \xi ^n\;d\mu (\xi ,\xi 
^{*})\;(\xi ^{*})^m$ with 
$d\mu (\xi ,\xi ^{*})=d^p\xi \;d^p\xi ^{*}\;e^{-\frac 12[\xi 
^{*},\xi ]}$ is
non-zero only when $n=m$. So in this integration, we identify
\begin{equation}
\xi ^n(\xi ^{*})^n\sim (n!)^2\sum_{i_1<\cdots <i_n}\xi 
_{i_1}\cdots \xi
_{i_n}\xi _{i_1}^{*}\cdots \xi _{i_n}^{*},
\end{equation}
\begin{equation}
\xi ^p(\xi ^{*})^p\sim (p!)^2\xi _1\cdots \xi _p\xi 
_1^{*}\cdots \xi _p^{*}.
\end{equation}
In the sum in (57) there are a total of ${p \choose n}$
terms\footnote{ ${p \choose n}$ denotes the ordinary binomial 
coefficient.}, and each
such term contributes only once in the paragrassman 
integration; thus,
\begin{equation}
\int \xi ^n\;d^p\xi \;d^p\xi ^{*}\;e^{-\frac 12[\xi ^{*},\xi 
]}\;(\xi
^{*})^n=(n!)^2 {p \choose n} =\frac{n!p!}{(p-n)!}=\{n\}!.
\end{equation}

\section{Proof of paragrassman differentiation formula:}

The left-differentiation with respect to $\xi $ is defined by
\begin{equation}
\frac \partial {\partial \xi }=\sum_{i=1}^p\;\frac \partial 
{\partial \xi
_i},
\end{equation}
where 
$$
\{\frac \partial {\partial \xi _i},\frac \partial {\partial 
\xi
_i}\}=0,\;\;[\frac \partial {\partial \xi _i},\frac \partial 
{\partial \xi
_j}]=0\;\;(i\neq j), 
$$
\begin{equation}
\{\frac \partial {\partial \xi _i},\xi _i\}=1,\;\;[\frac 
\partial {\partial
\xi _i},\xi _j]=0\;\;(i\neq j).
\end{equation}
In terms of Green components, $\xi ^n$ can be expressed as 
\begin{equation}
\xi ^n=n!\sum_{i_1<\cdots <i_n}\xi _{i_1}\cdots \xi _{i_n}
\end{equation}
where there are ${p \choose n}$ terms in the sum. For 
each $j$, in
\begin{equation}
\frac \partial {\partial \xi }\;\xi 
^n=n!\;\sum_{j=1}^p\;\frac \partial
{\partial \xi _j}(\sum_{i_1<\cdots <i_n}\xi _{i_1}\cdots \xi 
_{i_n}),
\end{equation}
there are ${ {p-1} \choose {n-1} }$ terms in the inner 
summation which involve 
$\xi _j$ and which survive after $\frac \partial {\partial 
\xi _j}$. So
there are a total of $n!\;p$ ${ {p-1} \choose {n-1} }$ terms 
on the 
``rhs'' of (63)
and they are of the form $\xi _{i_1}\cdots \xi _{i_{n-1}}$ 
with $i_1<\cdots
<i_{n-1}$. These are precisely the terms appearing in the 
Green component
expansion of $\xi ^{n-1}$. By considering the symmetry of the 
Green
components, we obtain the proportionality factor
\begin{equation}
\frac{n!p { {p-1} \choose {n-1} } }{(n-1)! { p \choose {n-1} 
} }=n(p+1-
n)=\{n\}
\end{equation}
so
\begin{equation}
\frac \partial {\partial \xi }\;\xi ^n=\{n\}\; \xi ^{n-
1}\;\;(0\leq n\leq p).
\end{equation}
Right-differentiation can be dealt with in a similar manner.

\end{document}